\documentclass[3p,twocolumn]{elsarticle}

\usepackage{graphicx}
\usepackage{multirow}
\begin{document}

\title{Tracing the evolution of nuclear forces under the similarity renormalization group}

\author{Calvin W. Johnson}
\address{Department of Physics, San Diego State University,
5500 Campanile Drive, San Diego, CA 92182-1233}
\ead{cjohnson@mail.sdsu.edu}
\tnotetext[t1]{This article is registered under preprint arXiv:1708:02601}


\begin{abstract}
I examine the evolution of nuclear forces under the similarity renormalization group (SRG)
using traces of the many-body configuration-space Hamiltonian.  While SRG is often said to ``soften'' the 
nuclear interaction, I provide numerical examples which paint a complementary point of view: 
  the primary effect of SRG, using the kinetic energy as the generator of the 
evolution, is to shift downward the diagonal 
matrix elements in the model space, while the off-diagonal elements  undergo 
significantly smaller changes.  By employing traces, I argue that this is a very 
natural outcome as one diagonalizes a matrix, and helps one to understand 
the success of SRG. 
\end{abstract}

\maketitle


\textit{Introduction}.
Nuclear structure theory has undergone a renaissance, driven by advances in high performance computing as well as by rigorous and systematic 
methodologies for  both \textit{ab initio}  nuclear forces, such as chiral effective field theory\cite{weinberg91,van1999effective,machleidt2011chiral} and for application of those forces to many-body calculations.
Included in the latter are the no-core shell model  \cite{navratil2000large,barrett2013ab}
 and the similarity renormalization group \cite{PhysRevD.48.5863,wegner1994flow,PhysRevC.75.061001,PhysRevLett.103.082501,bogner2010low}, which together have been very successful in calculating properties of light nuclei starting primarily from two-nucleon
 data.  It is often said that the similarity renormalization group (SRG) ``softens'' the nuclear interaction, improving convergence with model space size.  
  In this short paper I demonstrate that, at a gross level, the dominant effect of SRG on no-core shell model (NCSM) 
 calculations is to  shift low-lying energies down, with much smaller effects on wave functions. Using traces over the many-body model space, I argue that, in retrospect, the weakening of off-diagonal matrix elements and a much larger downward shift in diagonal elements go hand-in-hand.

The no-core shell model  is a configuration-interaction method, whereby one solves the 
nonrelativistic nuclear Schr\"odinger equation $\hat{H} | \Psi_i \rangle = E_i | \Psi_i \rangle$ as a matrix eigenvalue problem, typically in a basis of
Slater determinants, that is, antisymmetrized products of single-particle states in the lab frame.  
An important goal of modern nuclear structure theory is to carry out many-body 
calculations, in the NCSM or other methodologies, using  nucleon-nucleon forces fitted with high precision to experimental phase shifts \cite{stoks1994construction,wiringa1995accurate,entem2003accurate}.
These forces are generated in relative coordinates and then transformed to the lab frame via Talmi-Moshinsky-Brody brackets \cite{talmi1952tables,moshinsky1959transformation,brody1967tables}.   Now come two crucial concepts I  rely upon. The first is that finding eigenpairs 
involves a unitary transformation to a diagonal matrix. (Because one only wants low-lying eigenpairs and not all of them, one uses the Lanczos algorithm \cite{Whitehead}, but the 
basic idea remains.)  To diagonalize the ``full'' matrix 
is impractical, hence we must diagonalize in a smaller, truncated model space.  
Yet \textit{ab initio} nuclear forces   have large matrix elements connecting states of low and high relative momentum,
 which  historically and  
 phenomenologically was interpreted as a hard repulsive core.  In the shell model configuration basis this ``hard core'' becomes a strong coupling between 
 the truncated model space and the excluded space, driving the inclusion of many configurations to converge results as a function 
 of model space size, typically described by $N_\mathrm{max}$ (the number of excitations in an noninteracting harmonic oscillator space).

 In order to improve solutions in the model space, one turns to effective interaction theories. 
 Most of these are unitary or quasi-unitary 
 transformations, and one of the most widely applied is the similarity renormalization group, where one evolves a Hamiltonian by the differential equation
 \begin{equation}
 \frac{d}{ds} \hat{H}(s) = \left [ \hat{H}(s), \left [ \hat{H}(s), \hat{G} \right ] \right ]. \label{SRG}
 \end{equation}
 Here $\hat{G}$ is  the generator of the evolution and is often picked to be the kinetic energy $\hat{T}$.  If fully carried out, (\ref{SRG}) is a unitary transformation 
 of the Hamiltonian. (It also induces, however, many-body forces \cite{PhysRevLett.103.082501,PhysRevC.85.044003}, and as one typically carries out (\ref{SRG}) in just the two- or three-body systems, higher rank 
 forces are dropped and one has a loss of unitarity. A closely related methodology is the in-medium SRG\cite{tsukiyama2011medium,hergert2016medium}, which by normal ordering  approximately accounts
 for higher rank forces, although I do not consider it further.)
In other words, when a matrix is diagonal, each eigenstate is decoupled from all the rest, while  SRG and related techniques seek to improve 
the solution in the model space by approximately decoupling the model space from the excluded space.  In what follows I will gives examples of 
the effect of SRG on solutions in the model space, and then interpret those effects in terms of traces on the many-body Hamiltonian matrix.
 
\begin{figure}
\centering
\includegraphics[scale=0.33,clip]{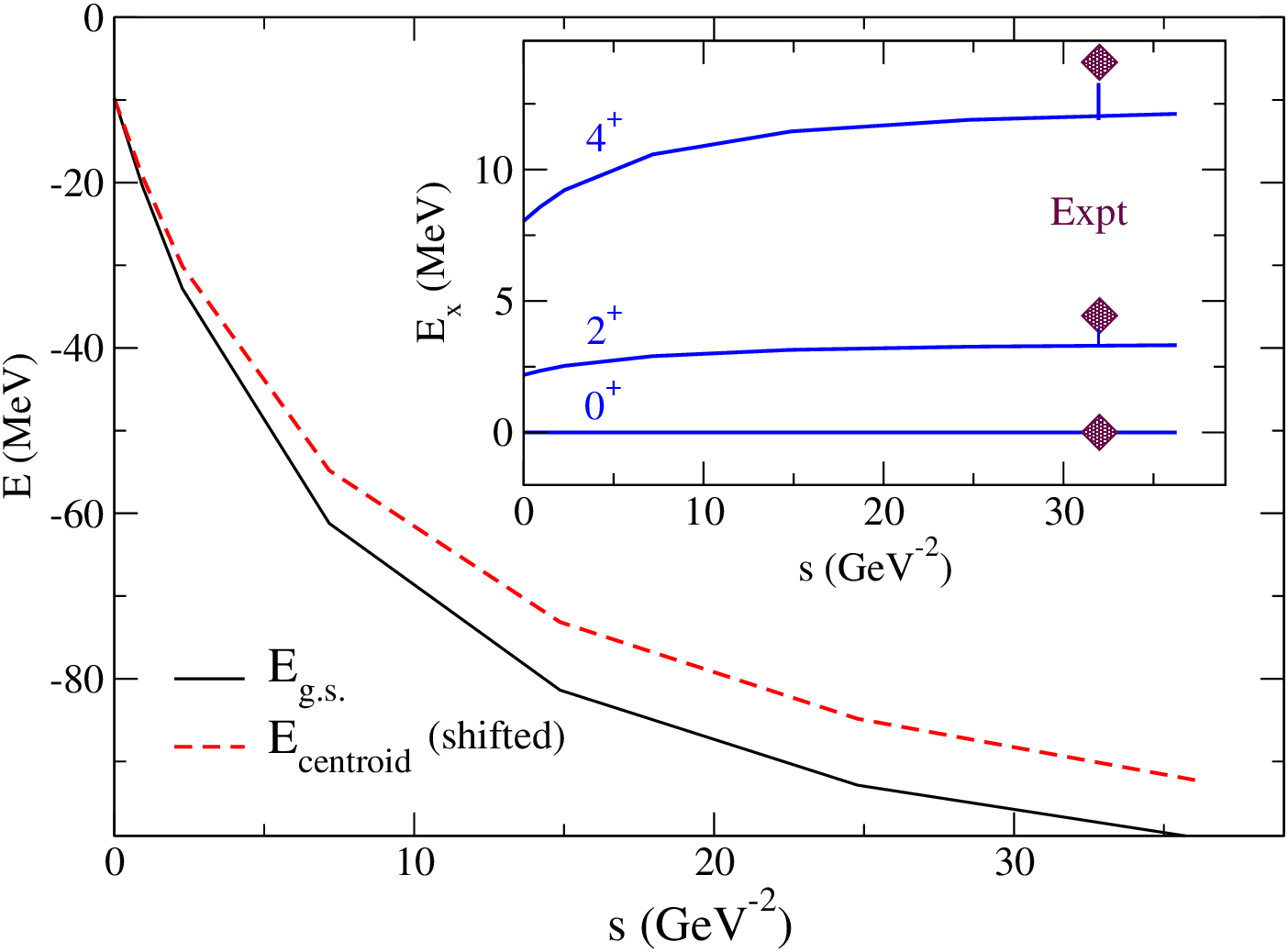}
\caption{(Color online) 
Evolution under the similarity renormlization group of ground state energy of $^{12}$C (black solid line) and excitation energies of the first $2^+$, $4^+$ states (inset), where the 
(maroon) diamonds show the experimental values for comparison.  The calculations were carried out in an $N_\mathrm{max}=8$ many-body space 
in a harmonic oscillator basis frequency of $\hbar \Omega = 20$ MeV, using the Entem-Machleidt chiral effective interaction evaluated to N3LO. 
Also shown
(red dashed line) is the evolution of the energy of the centroid, as defined in Eq.~(\ref{centroid_defn}); the centroid energy is shifted down by  221.88 MeV to show it tracking the ground state energy.}
\label{c12spect}
\end{figure}

 As an example of the NCSM using SRG-evolved forces, I consider $^{12}$C in a Slater determinant (lab frame) basis built from single-particle harmonic oscillator states with $\hbar \Omega = 20$ MeV,  allowing up to $N_\mathrm{max}=8$ excitations; such a space allows one to  exactly separate out spurious center-of-mass motion using the Palumbo-Glockle-Lawson method \cite{palumbo1967intrinsic,gloeckner1974spurious}.
 The interaction was derived in chiral effective field theory in next-to-next-to-next-to-leading order \cite{entem2003accurate} or N3LO; it was then evolved in 
 relative momentum space via SRG (with the kinetic energy as the generator $\hat{G}$ in Eq.~(\ref{SRG})) to various values of $s$, and then transformed to the lab frame via Talmi-Moshinsky-Brody brackets \cite{talmi1952tables,moshinsky1959transformation,brody1967tables}. 
(The representation of the interaction in relative coordinates has a finite cut-off; in harmonic oscillator states it is 
typically truncated at 100-300 $\hbar \Omega$ \cite{barrett2013ab}, which corresponds to $N_\mathrm{max}\sim 100$-$300$. 
In the lab frame, the non-spurious many-body states are  coupled to the ground state of
center-of-mass motion.  Thus  the non-spurious Hamiltonian in a many-body basis in the lab frame is also finite, though  $N_\mathrm{max}\sim 100$-$300$ is
much larger than any configuration-interaction code can fully handle, and so the Hamiltonian in single-particle coordinates, or the lab frame,
is always truncated.)
 Often one reinterprets the evolution in terms of a wavenumber cutoff, $\lambda = \frac{\sqrt{M_N}}{\hbar} s^{-1/4}$, which has units of fm$^{-1}$; here $M_N$ is the nucleon mass.  Although $\lambda$ suggests the ``resolution'' of the renormalized  interaction, 
 I retain $s$ to emphasize the evolutionary nature of SRG, so that $s=0$ is the original or unevolved interaction; this corresponds to $\lambda = \infty$. 
 The shell model calculations are carried out using the {\tt BIGSTICK} code \cite{BIGSTICK}.
 
 Fig.~1 shows how the ground state energy plunges as one goes from the bare force, $s=0$, to one evolved to $s$= 36.3 GeV$^{-2}$ which is equivalent to $\lambda = 2.0$ fm$^{-1}$. While the 
 ground state energy changes by nearly 90 MeV, the inset shows the excitation energies change hardly at all.   
Table \ref{gsdrop} shows $\Delta E_{g.s.} = E_{g.s.}(s=0) - E_{g.s.}(s= 36.3 \, \mathrm{GeV}^{-2})$, the change in the ground state energy under SRG, for several different $p$-shell nuclides, including 
 different $N_\mathrm{max}$ truncations.

 \begin{table}
 \caption{Changes in ground state energies and centroids for selected $p$-shell nuclides, including for different $N_\mathrm{max}$ truncations, under SRG evolution from $s=0$ ($\lambda = \infty$) to $s= 36.3$ GeV$^{-2}$ ($\lambda = 2.0$ fm$^{-1}$). Also shown is the 
 overlap, $| \langle \Psi_\mathrm{g.s.}(s= 36.3 \, \mathrm{GeV}^{-2} ) | \Psi_\mathrm{g.s.}(s=0) \rangle |^2$ of the ground state wave function vectors in configuration space.}
 \label{gsdrop}
 \begin{tabular}{|rcccc|}
 \hline
 Nuclide   & $N_\mathrm{max}$ & $\Delta E_{g.s.}$  & $\Delta E_\mathrm{centroid}$  & overlap \\
                &                               &  (MeV)                  &                  (MeV)  &  \\        
 \hline
 $^{6}$Li   &    8                         &  25.97                   &   16.81  &  0.948  \\
 $^{6}$Li   &    10                       &      21.54              &    14.37  &  0.956 \\
 $^{6}$Li   &    12                         &         17.40        &   12.46   & 0.960 \\
 \hline
 $^{7}$Li  &  8                         &           33.58                  &    24.13            &  0.940 \\
 $^{7}$Li  &  10                         &            28.182                 &      21.03        &  0.950 \\
\hline
  $^{8}$Be  &  8                         &              46.33               &      33.36        &  0.927 \\
  $^{8}$Be  &  10                         &            39.48                 &   29.52         & 0.937  \\
\hline
  $^{9}$Be  &  8                         &               52.96              &  43.45     &        0.915  \\
  $^{9}$Be  &  10                         &            45.33                 &      38.93   &  0.927     \\
\hline
  $^{10}$B  &  8                         &           62.88                  &       55.40     &  0.896  \\
\hline
  $^{12}$C  &  8                         &              89.55               &       82.65     &  0.860  \\
 \hline
 \end{tabular}
 \end{table}

 One can take this further; in the {\tt BIGSTICK} code, as in 
 all shell-model diagonalization codes,  wave functions are represented as vectors, whose components are amplitudes in  the basis of Slater determinants.  Formally, 
 under the unitary transformation induced by SRG, the interpretation of that basis evolves along with the Hamiltonian.   But Table \ref{gsdrop}, which also 
 gives the overlap  $| \langle \Psi_\mathrm{g.s.}(s= 36.3 \, \mathrm{GeV}^{-2} ) | \Psi_\mathrm{g.s.}(s=0) \rangle |^2$, 
 shows that the vectors change only a small amount.
Of course, even small changes in the wave function can lead to large changes in specific matrix elements, particularly if there are large cancellations.
But at a gross level,  \textit{the largest effect of SRG evolution is 
  to shift down the low-lying energies}, with    smaller changes to the wave functions in occupation space and to excitation energies.  
 

 While this is gratifying, and perhaps what we would most want from any effective theory, is there someway we can understand this?

\textit{Spectral distribution theory}.  To analyze what is happening under SRG, I turn to spectral distribution theory (SDT), sometimes also called statistical spectroscopy 
\cite{PhysRevC.85.044003,french1967measures,french1983statistical,wong1986nuclear}.  
The key idea of SDT is the use of the average over an $N$-dimensional many-body space ${\cal S} = \{ | i \rangle \}$:
\begin{equation}
\langle \hat{R} \rangle^{({\cal S})} \equiv 
 \frac{1}{N} \sum_{i \in {\cal S}} \langle i | \hat{R} | i \rangle 
\label{trace}
\end{equation}
Note that this is \textit{not} an expectation value; following practioners  \cite{french1983statistical,wong1986nuclear,launey2014program}, 
 a superscript $({\cal S})$ emphasizes this difference.
 
 Traces are a powerful tool. By taking the trace of Eq.~(\ref{SRG}), and using the cyclic property of traces, one finds almost trivially that 
 $\frac{d}{ds} \mathrm{ tr}\, \hat{H}(s) = 0$, as well as  $\frac{d}{ds} \mathrm{ tr}\, \hat{H^2}(s) = 0$ and higher moments, proving that 
 SRG, if carried out exactly, is a unitary transformation. 

Remember, however, that in the version of SRG discussed here the Hamiltonian is evolved in relative coordinates and 
and when transformed to single-particle, lab-frame coordinates is very large and must be truncated.
 Thus, most calculations are carried out in a truncated model space ${\cal S}$, where the trace is not preserved. (The situation 
 for in-medium SRG is different and not discussed here.) 
   Nonetheless, 
one can consider the \textit{centroid}, the average of the many-body Hamiltonian in ${\cal S}$ :
\begin{equation}
E_\mathrm{centroid} = \langle \hat{H} \rangle^{({\cal S})}. \label{centroid_defn}
\end{equation}
Table \ref{gsdrop} shows the change $\Delta {E}_\mathrm{centroid} = E_\mathrm{centroid}(s=0) - {E}_\mathrm{centroid}(s=36.3 \, \mathrm{GeV}^{-2} )$ for a selection of $p$-shell nuclides. Note that a significant fraction of the drop of the ground state energy 
comes from the shift in the centroid. This is shown in detail for $^{12}$C in Fig.~1, where the evolution of the centroid with $s$ tracks the evolution of the ground state energy;
in that figure I've shifted the centroid down by 221.9 MeV so the tracking is more clear.  In other words, a significant effect of SRG evolution is 
not only to shift downwards the low-lying eigenstates, but to shift \textit{all} the states in the model space downwards.

One of the key tools of carrying out traces in SDT is to rewrite in terms of number operators.  With that in mind, the $N_\mathrm{max}=8$ data in Table \ref{gsdrop} has
$\Delta {E}_\mathrm{centroid} \approx 0.6 A(A-1)$ MeV.  There is a small but nontrivial additional dependence on $T_z$.   Furthermore, 
when scaled by $A(A-1)$, the evolution in $s$ of the centroids for fixed $N_\mathrm{max}$,  shown in Fig.~\ref{c12spect} for $^{12}$C (dashed line), all fall on the same curve.  SDT expects this: there is a operator $\hat{N}(\hat{N}-1)$ and its coefficient evolves with $s$. Although this `universal' curve does not have an obvious analytic form,  one might calculate directly the evolution 
of the centroid in terms of number operators, including higher order terms.Note that while the bulk of the 
change in the ground state energy comes from the shift in the centroid, the remainder is non-trivial and is \textit{not} `universal.'

The centroid is just the average of the diagonal elements in the model space.  One can track the evolution in more detail by using a finer tool, \textit{configuration centroids}. 
A \textit{configuration} is 
the occupancy of different orbits, for 
example: $(0s_{1/2})^2 (0p_{3/2})^2$, $(0s_{1/2})^2 (0p_{3/2})^1, (0p_{1/2})^1$, etc.. Then one can define a subspace, call the \textit{configuration partition} 
but sometimes just the configuration, which is  the set of all  states 
described by the same orbital occupancies. 
It turns out to be easy to compute the trace of the Hamiltonian within any configuration partition so defined \cite{wong1986nuclear}. If we label configurations by $\alpha$ with a projection operator 
$\hat{P}_\alpha = \sum_{i \in \alpha} | i \rangle \langle i |$, then the configuration dimension is just $N_\alpha = \mathrm{tr} \, \hat{P}_\alpha$ and the configuration centroid is 
$\bar{E}_\alpha = N_\alpha^{-1} \mathrm{tr} \, \hat{P}_\alpha \hat{H}$.  By subdividing the trace into configuration partitions, we can use configuration centroids to follow the evolution of the 
diagonal matrix elements. 

 \begin{figure}
\centering
\includegraphics[scale=0.33,clip]{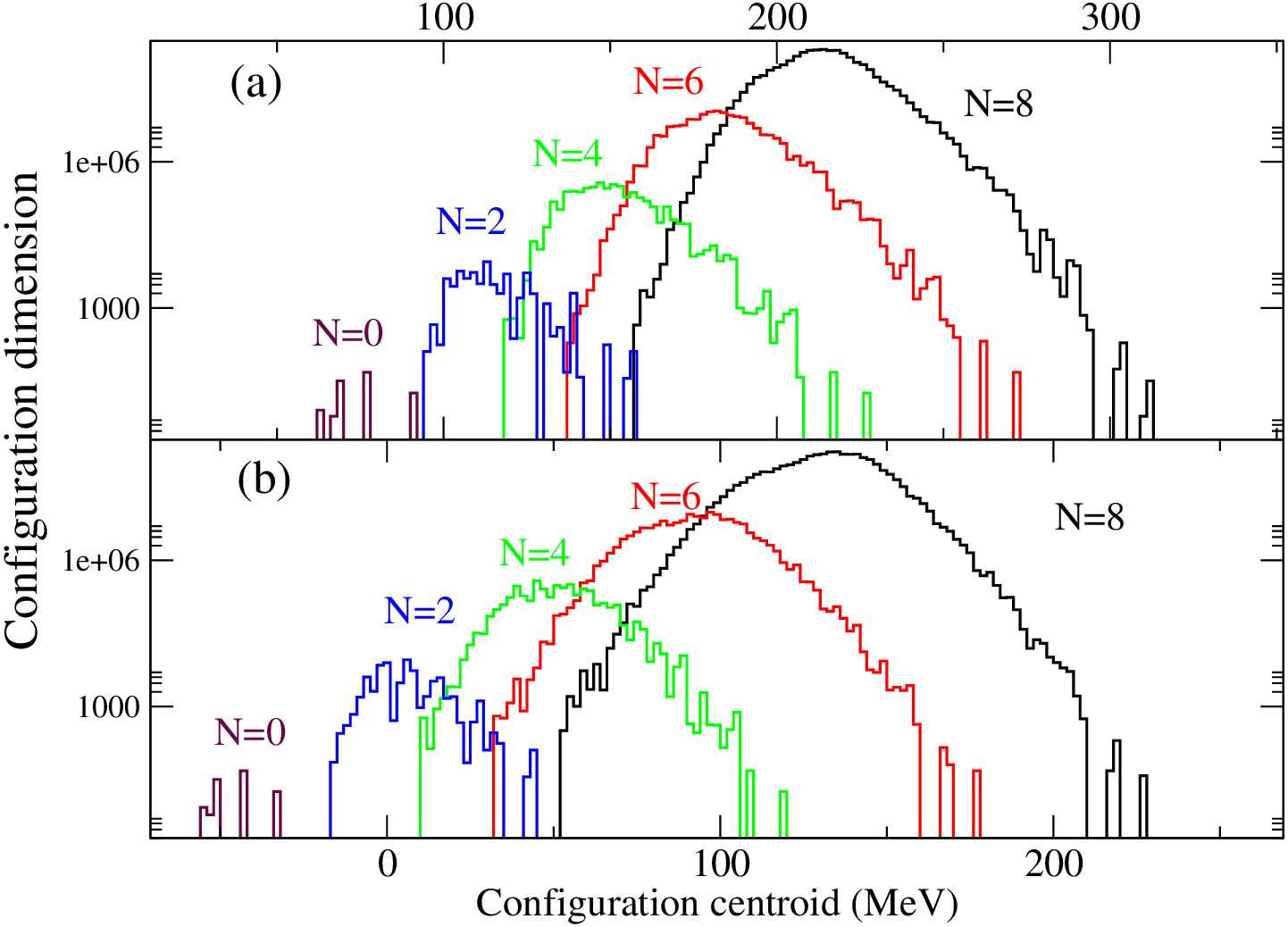}
\caption{Evolution of configuration centroids for $^{12}$C under the similarity renormalization group, from $s=0$, or $\lambda=\infty$ (top panel) to $s=36.3 \, \mathrm{GeV}^{-2}$ , or $\lambda = 2.0 \, \mathrm{fm}^{-1}$ (bottom panel). The abscissa is the configuration centroid, 
while the ordinate is the dimension of each configuration subspace.  The two plots are shifted so their 
respective total centroids (\ref{centroid_defn}) are aligned.  
Here $N$ is the \textit{fixed} number of excitations in harmonic oscillator many-body space, so that $N=4$ includes only 
4 $\hbar \Omega$ excitation but no $0\hbar \Omega$, no $2 \hbar \Omega$, etc. .}
\label{allcentroids}
\end{figure}

Fig.~\ref{allcentroids} plots the distribution of configurations centroids for  $^{12}$C, both  unevolved ($s=0$, top panel) and highly evolved (bottom panel) Hamiltonians, with the centroids combined into bins of width 2 MeV.  The $y$-axis is the dimension of the binned configuration subspaces, on a log plot.
The total centroid for each plot are aligned. Here $N$ is the number of excitations in the harmonic oscillator space.  
To be clear, here $N$ is fixed, not an upper limit; while CI calculations labeled by  $N_\mathrm{max}$ include states with $N=N_\mathrm{max}, 
N_\mathrm{max}-2, N_\mathrm{max}-4,\ldots$ ($\Delta N=2$ preserves parity), each  curve in Fig.~\ref{allcentroids}
is for all configurations of a single fixed $N$.
Despite an overall shift of about 82 MeV 
in the centroids, the distribution of configuration centroids is remarkably unchanged. The main visible difference is the evolved configuration centroids are ``stretched out.'' 
 
 What about off-diagonal elements?  Here I exploit another idea from SDT, that by using traces one can establish an inner product 
 on operators \cite{french1967measures}:
 \begin{eqnarray}
\left (\hat{H}_1, \hat{H}_2  \right) \equiv \nonumber \\
 \left \langle \left (  \hat{H}_1 -    \langle \hat{H}_1 \rangle^{({\cal S})}  \right) 
\left (  \hat{H}_2- 
 \langle \hat{H}_2 \rangle^{({\cal S})}   \right) \right \rangle  \label{SDTmetric} \\
=  \langle \hat{H}_1 \hat{H}_2 \rangle^{({\cal S})}   - 
\langle \hat{H}_1 \rangle^{({\cal S})}   \,\,\,
 \langle \hat{H}_2\rangle^{({\cal S})} . \nonumber
\end{eqnarray}
With an inner product so defined, one has a metric on the space of Hamiltonians and can 
compare how close or distant two Hamiltonians are; the magnitude of a Hamiltonian, $\| \hat{H} \| = \sqrt{( \hat{H}, \hat{H} )}$, is its width, and  the ``angle'' between two Hamiltonians by 
$\cos \theta = (\hat{H}_1, \hat{H}_2) \|\hat{H}_1 \|^{-1} \|\hat{H}_2 \|^{-1}$. 
Note that this definition of the metric is independent of the centroids, which is sensible as 
centroids affect only absolute energies and not excitation energies nor wave functions.  

 Using Eq.~(\ref{SRG}) and 
the cyclic property of traces,
\begin{equation}
\frac{d}{ds} \mathrm{tr} \,  \hat{H}(s) \hat{G}  = \mathrm{tr} \,  [ \hat{H}(s), \hat{G} ]^\dagger [ \hat{H}(s), \hat{G} ]. \label{SRGoverlap}
\end{equation}
 But the righthand side is the trace of an operator of the form $\hat{A}^\dagger \hat{A}$, which  manifestly has real and nonnegative eigenvalues
 and  a real, nonnegative trace. Thus,  under generic SRG,  $\mathrm{tr} \,    \hat{H}(s) \hat{G}  $ can only increase; as the magnitudes of $\hat{H}(s)$ and, trivially, 
 $\hat{G}$ are invariant, this means the ``angle'' defined by the inner product (\ref{SDTmetric}) between them can only get smaller and they become more and more parallel. Furthermore, this 
 evolution stops  when $ [ \hat{H}(s), \hat{G} ]=0$, that is, the evolved Hamiltonian  commutes with the generator. 
Thus SRG  drives a Hamiltonian ``towards'' its generator.
 $\hat{H}(s)$ cannot 
 become proportional to $\hat{G}$, as the eigenvalues, invariant under a unitary transformation, are different, but if one lets $s \rightarrow \infty$ they will have the same 
 eigenvectors. 
 
 \textit{Interpretation using traces in the model space}.
 Now let's discuss 
 these empirical results--the large, coherent shifts in low-lying energies, and the relatively modest changes in the wave function vectors--through the lens of 
 spectral distribution theory, using traces as a fundamental tool to investigate what happens under the unitary transformation induced by SRG.   
 To do so, 
 one must pay  attention to is the difference between  
taking traces in the full space and in the much smaller model space.

 Let $\hat{H}$ be the original Hamiltonian in the full space, and let $\hat{H}^\prime = \hat{U}^\dagger \hat{H} \hat{U}$ be the transformed
Hamiltonian.  Since the full space is too large to work in, let the much smaller model space be
${\cal S}$ with a projection operator $\hat{P}_{\cal S}$. 

The trace of any matrix (and of any power of that matrix) is preserved under unitary transformations:
 $ \mathrm{tr}\, \hat{H} = \mathrm{tr} \, \hat{U}^\dagger \hat{H} \hat{U} =\mathrm{tr}\, \hat{H}^\prime$, and similarly  $ \mathrm{tr}\, \hat{H}^2 =   \mathrm{tr}\, (\hat{H} ^\prime)^2$.   
Let's divide up the contributions to the trace of $\hat{H}^2$ into diagonal and off-diagonal pieces:
$\mathrm{tr}\, \hat{H}^2 = \sum_i H_{ii}^2 + \sum_{i \neq j} H_{ij}^2$. If after a unitary transformation the matrix 
$\hat{H}^\prime$ is diagonal, then 
$\mathrm{tr}\, (\hat{H}^\prime)^2$ has no off-diagonal contributions. Thus 
$\mathrm{tr}\, \hat{H}^2 =\mathrm{tr}\, \hat{H}^{\prime 2} = \sum_i (H^{\prime})^2_{ii} \geq \sum_i H_{ii}^2  $. 
That is, on average, the magnitude of the diagonal elements of the transformed matrix $\hat{H}^\prime$ must be larger than 
those of the original matrix $\hat{H}$.  

Furthermore,  the change in the diagonal matrix elements will generally be much 
larger than that of the off-diagonal matrix elements. How so? Even for sparse matrices, there will be many more 
off-diagonal matrix elements than diagonal.  Suppose in an $N$-dimensional space there are $\sim N \times M$ non-zero off-diagonal matrix elements, 
so that $M/N \sim$ the sparsity. Further suppose the off-diagonal matrix elements all have roughly the same 
magnitude: call it $\gamma$.   On average for every diagonal matrix element there are $\sim M$ 
non-zero matrix elements, and  $\mathrm{tr}\, \hat{H}^2  \sim \sum_i H_{ii}^2 + NM \gamma^2$. 
Then on average $(H^\prime_{ii})^2 \sim (H_{ii})^2 + M \gamma^2$ so that the root-mean-square change in the 
diagonal matrix elements,
$\delta H_{ii}  \equiv \sqrt{ (H^\prime_{ii})^2 - (H_{ii})^2} \sim \sqrt{M} \, \gamma.$
Even for tiny sparsities,  $M$ will be a large number. Hence the average changes of diagonal matrix elements will be
substantially larger than the changes to the off-diagonal matrix elements.  

\begin{figure}
\centering
\includegraphics[scale=0.33,clip]{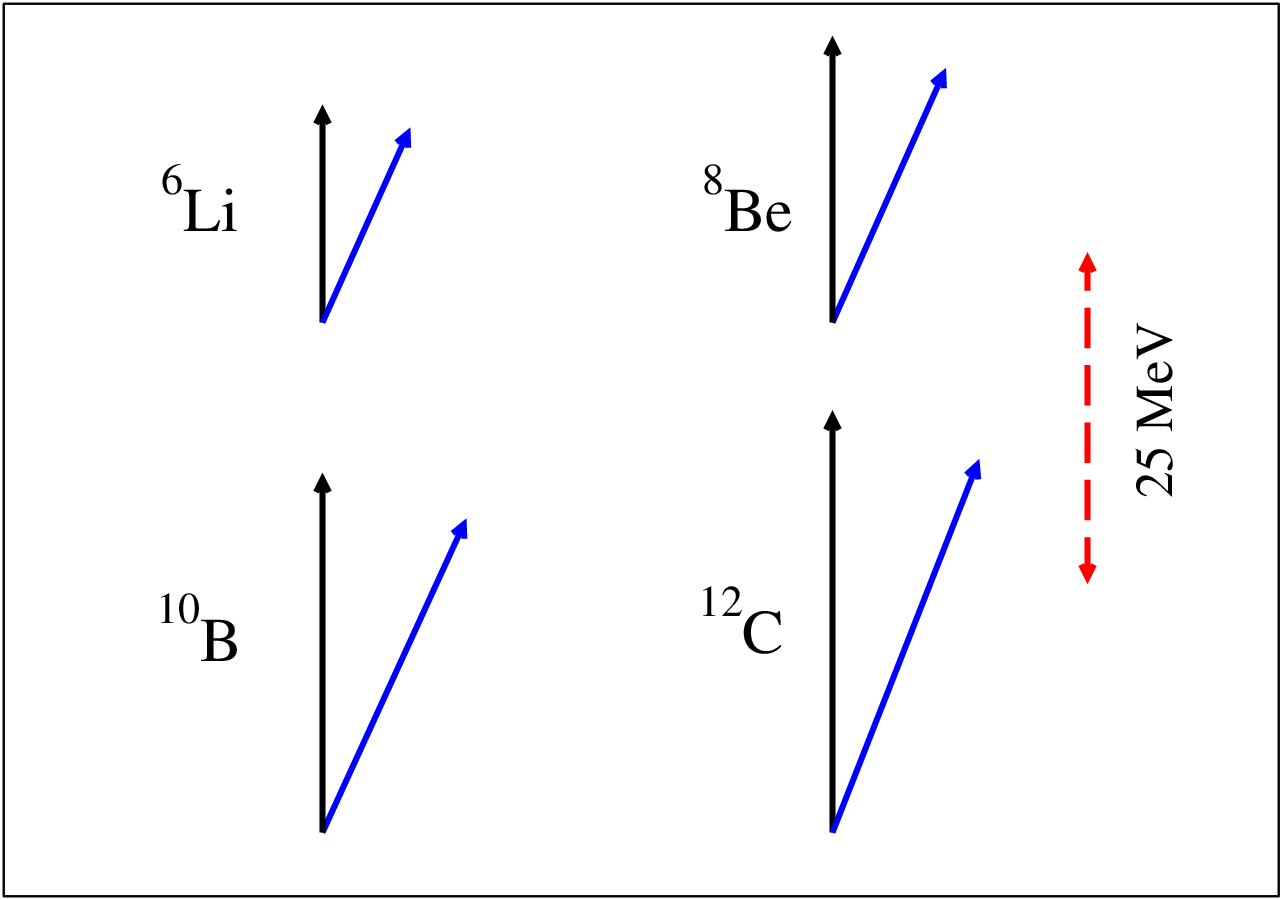}
\caption{A representation of the evolution of interactions under SRG, as measured by the spectral distribution theory 
inner product, Eq.~(\ref{SDTmetric}), for selected nuclides.  The black, vertical arrows represent the unevolved interaction, at 
$s=0$ or $\lambda = \infty$, while the blue, tilted arrows are the interactions evolved to $s=36.3 \, \mathrm{GeV}^{-2}$ or 
$\lambda = 2.0 \, \mathrm{fm}^{-1}$. Here the evolved interaction is defined as $\hat{V}(s) = \hat{H}(s) - \hat{T}$, where $\hat{T}$ is the 
kinetic energy. 
The red dashed line gives the scale.  Although not shown, results for $A=2,3$ are very similar, although with smaller absolute magnitudes.}
\label{hamdot}
\end{figure}

This argument suggests that despite the large shift 
in the centroid in the model space, the changes to the off-diagonal matrix elements are much smaller.  I argue  this is true even when one is
not fully diagonalizing but only applying a unitary transformation which ``softens'' the interaction.
As further evidence,  I use 
 the metric introduced in Eq.~(\ref{SDTmetric}).  Fig.~\ref{hamdot} graphically compares  
two Hamiltonian as the two vectors, with the magnitudes of each operator and the angle between them defined by 
(\ref{SDTmetric}),   for four $p$-shell nuclides.  I considered  isospin zero many-body states; the results 
depend only weakly upon isospin.  To focus on the actual evolution of the operators, I consider only the ``interaction,'' defined as 
$\hat{V}(s) = \hat{H}(s) - \hat{T}$, although such an interaction has components evolved from the original kinetic energy. 
The inner products were calculated using a publically available code  \cite{launey2014program}; because this code does not allow for 
$N_\mathrm{max}$ truncations, I instead truncated on the number of harmonic oscillator shells.   I show the results for 7 major harmonic oscillator 
shells, that is, for maximum principal quantum number $N=6$, or up through the $3s$-$2d$-$1g$-$0i$ shell.  The results, however, do not change 
very much as one goes from 5 to 7 major harmonic oscillator shells, and one can understand this as the higher excitations being nearly completely 
dominated by kinetic energy \cite{PhysRevC.94.064320}.

Fig.~\ref{hamdot} shows the change in the interactions is modest.  (Furthermore, although not included in Fig.~\ref{hamdot}, the angle between evolved and unevolved interactions in this space for $A=2,3$ is 
nearly identical to the cases shown.)
 This is in concordance with the previous empirical observation that under SRG evolution the change of the wave functions, as 
vectors in an abstract space of Slater determinants, is modest.  Let me emphasize, however, that changes to the 
off-diagonal matrix elements are not negligible. If I approximate the SRG transformation by taking the unevolved Hamiltonian and simply swapping in 
the evolved configuration centroids, the shift in the ground state energy is overestimated by several MeV.

Eq.~(\ref{SRGoverlap}) proves that SRG evolves a Hamiltonian towards the generator. That proof, however, only applies in the full space. 
In the truncated model space, I find little change in the angle between the interaction and the kinetic energy, and in fact the angle increases, though 
by less than a degree.

Because the ``full'' Hamiltonian in the lab frame is generally far too large to be tractable, it must be truncated, which leads to 
loss of unitarity. 
But truncation is not the only cause of loss of unitarity.   Three-body and higher-rank forces,  induced by SRG \cite{PhysRevLett.103.082501,PhysRevC.85.044003} 
also  contribute to unitarity of the transformation.  Previous work \cite{PhysRevC.76.034302} as well as the success of 
the in-medium SRG \cite{tsukiyama2011medium,hergert2016medium} has suggested the most important contributions of these many-body forces 
are those written as density- or state-dependent operators, especially those which contribute to the monopole terms \cite{PhysRevC.76.034302}, that is, the centroid and the configuration centroids. This is completely consistent with the results presented here.

\textit{Conclusions.}   Using traces or averages of operators in a many-body space, also called spectral distribution theory,  is a powerful tool for understanding the evolution of nuclear forces under
the similarity renormalization group.   The primary effect of SRG is to simply to shift the model space centroid  downwards. In contrast configuration centroids, or averages within occupancy-defined subspaces,are changed only a little relative to the total centroid.  The inner product in SDT also allows one 
to look at off-diagonal matrix elements, which change only a modest amount under SRG.
This picture--large, coherent shifts in the diagonal matrix elements and relatively much smaller changes to off-diagonal matrix elements--arises 
naturally when diagonalizing a Hamiltonian and is complementary to the  conception of  SRG ``softening'' the nuclear interaction.
Finally, as the centroid and configuration centroids are written in terms of number operators, and the shift in the centroid  
is proportional to $A(A-1)$, I suggest one might compute 
separately the evolution of the centroid, not only as a function of $A(A-1)$ but also of higher powers of the number operator. Induced 
higher-particle-rank forces are important, and  pieces proportional to powers of number operators account for the bulk of the effect \cite{PhysRevC.76.034302}.

I thank P. Navr\v{a}til for his code to generate the Entem-Machleidt N3LO force and apply SRG to it, and K.~D.~Launey for helpful discussions. 
This material is based upon work supported by the U.S. Department of Energy, Office of Science, Office of Nuclear Physics, 
under Award Number  DE-FG02-03ER41272. This research used resources of the Argonne Leadership Computing Facility, which is a DOE Office of Science User Facility supported under Contract DE-AC02-06CH11357, and of the National Energy Research Scientific Computing Center, a DOE Office of Science User Facility supported by the Office of Science of the U.S. Department of Energy under Contract No. DE-AC02-05CH11231.

\bibliographystyle{apsrev4-1}
\bibliography{johnsonmaster}
\end{document}